\documentclass[a4paper,11pt]{article}
\usepackage{pos}

\title{Recent Results from VERITAS AGN Observations}
 \ShortTitle{Recent Results from VERITAS AGN Observations}

\author*[a]{Wystan Benbow}
\affiliation[a]{Center for Astrophysics | Harvard \& Smithsonian,\\ 60 Garden St, Cambridge, MA, 02138, USA\\}

\forColl{VERITAS}

\emailAdd{wbenbow@cfa.harvard.edu}

\abstract{VERITAS is one of the world’s most sensitive detectors of astrophysical very high energy (VHE; E $>$ 100 GeV) gamma rays. This observatory has operated for $\sim$14 years, and nearly 7,000 hours of its observations have been targeted on active galactic nuclei (AGN). Approximately 300 AGN were observed with VERITAS, and 40 are detected. These studies are generally accompanied by contemporaneous, broadband observations, which enable detailed probes of the underlying jet-powered processes. Recent scientific results from VERITAS AGN observations are presented. }

\FullConference{37$^{\rm{th}}$ International Cosmic Ray Conference (ICRC 2021)\\
		July 12th -- 23rd, 2021\\
		Online -- Berlin, Germany}

\begin{document}
\maketitle

\section{Introduction}

AGN comprise $\sim$34\% of the VHE sky catalog and are the 
most numerous class of identified VHE $\gamma$-ray source \cite{TEVCAT}.
These objects have long dominated the observing programs of Northern VHE observatories
such as VERITAS \cite{VERITAS_spec}. As of \emph{ICRC 2021}, AGN
comprise approximately 52\% of the known VHE sources 
with Northern declinations ($\delta > 0^{\circ}$), and 76\%
of the 84 VHE-detected AGN are located in the North.
These dominantly non-thermal sources have
spectral energy distributions (SEDs) spanning from radio waves
through $\gamma$-rays. Their emission is highly
variable at all wavelengths, and on time scales as short as minutes.
This leads to an emphasis on using contemporaneous multi-wavelength (MWL) 
observation campaigns to probe their underlying physics. It is generally agreed that 
the VHE $\gamma$-ray emission from AGN is produced in their relativistic jet, in a compact region
near their central, supermassive black hole. However, some recent detections
show indications for VHE emission produced on larger and/or more distant scales
(see, e.g., \cite{VERITAS_1441}). The characteristic double-peaked SEDs observed from
AGN naturally suggest emission models where the low-energy peak is the 
synchrotron radiation from a population of relativistic electrons in the relativistic jet, and the
high-energy photons are the products of inverse-Compton scattering from this population
(e.g. a synchrotron self-Compton (SSC) model as in \cite{VERITAS_0229}). While VHE emission models dominated by
leptonic particle-acceleration processes in the accretion-powered jets remain favored,
other models remain viable (see, e.g., \cite{PKS1424_model} and the discussion / references therein). 

Blazars are AGN whose relativisitc jets are pointed along the line of sight
towards Earth.  They form the dominant class ($\sim$95\%) of VHE AGN. 
Of the 78 VHE blazars, 67 ($\sim$90\%) are BL Lac objects,
8 are Flat Spectrum Radio Quasars (FSRQs) and 3 have uncertain sub-classification.
Of the remaining 6 VHE AGN, at least four are nearby ($z < 0.022$) FR-I radio galaxies, 
and the radio-galaxy / blazar classification for another 2 remains debated. In any case, 
the jets for these 6 objects are not strongly misaligned. The sub-classification of 
BL Lac objects, based on the location of their lower-energy SED peak, is important.
Within the VHE catalog, 53 BL Lacs are high-frequency-peaked (HBLs), 10 are intermediate-frequency-peaked
(IBLs), 2 are low-frequency-peaked (LBLs) and two are unclassified. The VHE blazar catalog currently covers
a redshift range from $z = 0.030$ to $z = 0.954$, but at least
$\sim$20\% of the objects have unknown redshift and many catalog values are uncertain.  
This is in large part due to the absence of optical features in the spectra of BL Lac objects.
Generally, the VHE AGN catalog is peaked at nearby redshifts (e.g., $\sim$55\% have $z < 0.2$)
but $\sim$10\% of the VHE objects have $z >0.5$ (primarily FSRQs). Aside from energetics
considerations, the major contributor to this redshift distribution is 
attenuation of VHE photons in a distance- and energy-dependent manner by the
extragalactic background light (EBL) \cite{VERITAS_EBL}.

In general, the observational goals of the VERITAS AGN (radio galaxy and
blazar) Program are to discover new VHE AGN, and to make precision measurements 
of VHE AGN spectra and their variability patterns.  There are two empirical
qualities of the AGN population that drive the strategy behind the VERITAS
program.  The first is that most VHE AGN have shown at 
least some VHE flux variability. Indeed major outbursts
are perhaps one of the defining characteristics of the VHE field 
(see, e.g., \cite{VERITAS_421_flare}), and roughly one-third of VHE AGN 
are only detected during brief flares.  These flare-only AGN include
most of the non-HBL blazars.  While extreme episodes may define the field, these
rapid (minute-scale), large-scale (factor of 100) flux
variations are very rare.  Most VHE flux variations are mild (factor of 2-3)
and many have comparatively long time scales (e.g., an observing season).
In general the time-scale and strength of the variations observed 
depends on the average VHE flux, but this may have biases (e.g., instrument
sensitivity, target selection, etc.) The VERITAS AGN
Program therefore attempts to identify and follow-up on VHE AGN
flares, guided in part by brighter than average VHE flux.
The other important empirical quality is that the observed photon spectra of
VHE AGN are often soft ($\Gamma_{avg} \sim 3.5$), and very few VHE
AGN are detected above 1 TeV. This partly due to the high-energy SED peak 
location, which may depend on the underlying physics,
and partly due to propagation effects (e.g., EBL absorption).
AGN with hard VHE spectra are particularly interesting, and the 
VERITAS AGN Program also focuses on hard spectrum VHE blazars 
and generating statistics above above 1 TeV.  A study of the spectra measured by VERITAS from
these VHE blazars can be found in these proceedings \cite{Feng_ICRC21}.

The VERITAS AGN Program strongly leverages contemporaneous MWL observations from numerous ground-
and space-based facilities. Swift XRT/UVOT, \emph{Fermi}-LAT, the FLWO 48'' optical telescope form the core
of this regular lower-energy coverage, which exists for all Northern VHE AGN.  
After an initial focus
on expanding the VHE AGN catalog from 2007-10, the VERITAS AGN program has since emphasized studies
of the known VHE AGN population with a goal of identifying and intensely
observing major flares.  The target list has evolved but the general method
was to sample each selected VHE AGN on a regular cadence, thereby building high-statistics data sets while
searching for flares. 
VERITAS AGN observations are timed to the MWL coverage to enable 
fully-constrained modeling of each VHE AGN's SED (see, e.g., \cite{VERITAS_0229}).
From 2013-2018, the program sampled each Northern VHE AGN ($\sim$56 targets), but
it was recently streamlined to focus on more intense studies of select ($\sim$23) targets. 
Our current focus is on the most variable VHE sources (i.e. IBLs), bright HBLs, and hard-spectrum HBLs.
This program has yielded detailed, decade-long MWL light curves for
numerous ($\sim$20) sources and significant coverage for the entire Northern population,
in addition to the identification of numerous VHE flares.  The ongoing
sampling continues to improve these efforts, and the various long-term MWL light 
curves will enable flux and spectral correlation studies that may indicate
commonalities in the origin of each AGN's emission. The large ensemble of 
precision VHE AGN spectra will also improve and has already proved
useful for generating a variety of cosmological measurements
such as constraints on the the density of the EBL \cite{VERITAS_EBL}
and the strength of the intergalactic magnetic field (IGMF) \cite{VERITAS_IGMF}.

\begin{figure*}[t]
   \centerline{ {\includegraphics[width=3.75in]{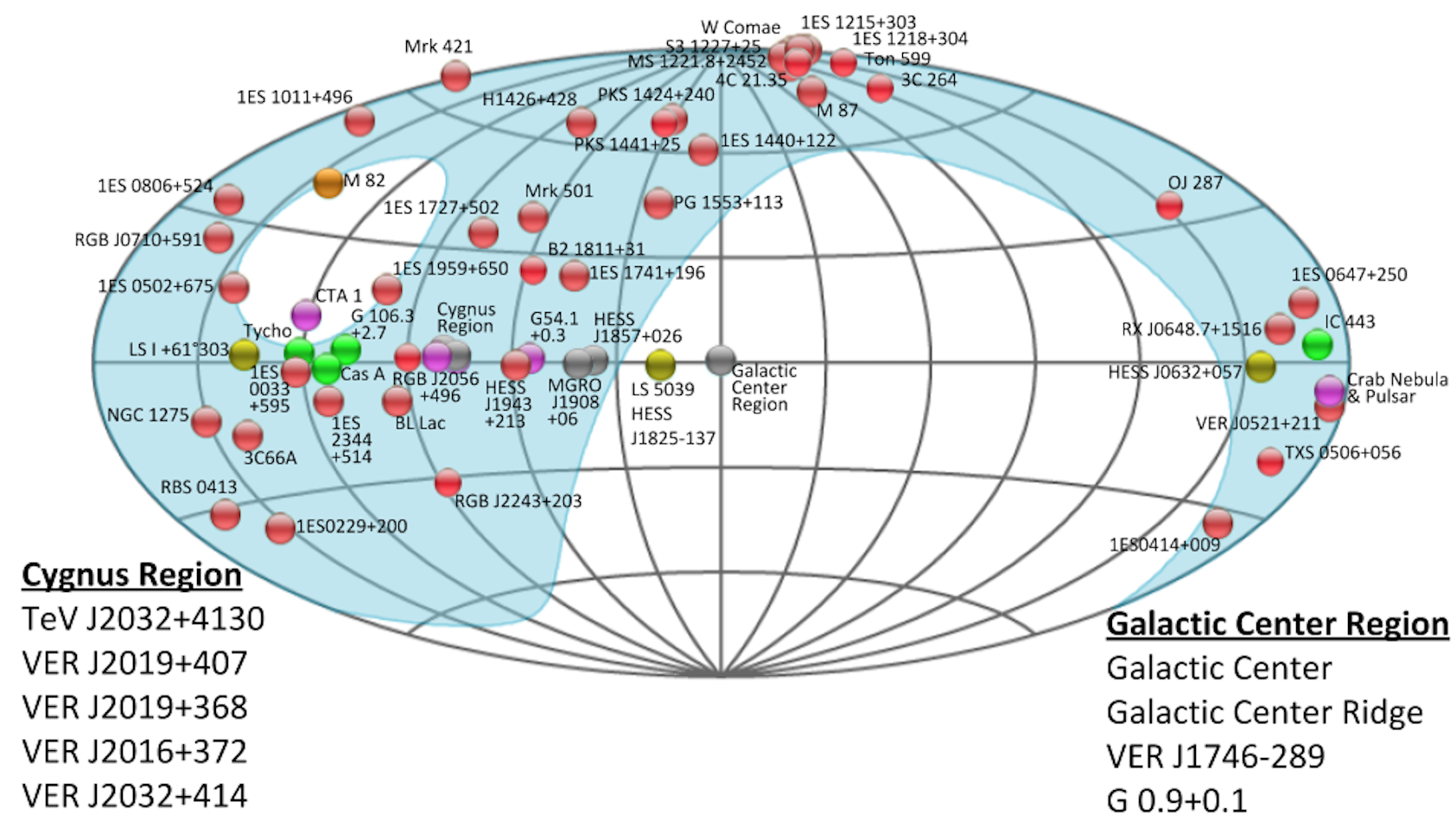} } }
\vspace{-0.3cm}
   \caption{{\footnotesize The VERITAS catalog in Galactic
       coordinates. The red circles show the 40 AGN detected using VERITAS; other astrophysical classes are shown with
       different colors. The blue region is visible to VERITAS at
       zenith angles $<$35$^{\circ}$. Most of the VERITAS AGN catalog is given in \cite{Benbow19}; only the new IBL B2\,1811+31 (see below) is missing.}}
   \label{VERITAS_catalog}
\vspace{-0.3cm}
 \end{figure*}

\section{VERITAS AGN Program}

VERITAS began full-scale operations in 2007 at the F. L. Whipple Observatory 
in Arizona, USA.  After fourteen years, it remains the world's most sensitive observatory between $\sim$85 GeV
and $\sim$30 TeV. VERITAS scientists use the array of four Cherenkov telescopes
to study the Northern sky during $\sim$10-month, monsoon-limited seasons (September $-$ July).
The present sensitivity was achieved following a series of
upgrades in 2012, and VERITAS can detect an object with $\sim$1\% Crab Nebula 
flux (1\% Crab) in less than 25 hours. Photon spectra can be generated above
$\sim$100 GeV with typical systematic errors of $\sim$0.1 on the photon index
($\Gamma$) and $\sim$20\% on the flux.

The VERITAS collaboration has acquired $\sim$16,300 h of observations. 
While successful, the past two seasons were challenging
due to the global pandemic. The 2019-20 season ended $\sim$4 months
early causing an estimated loss of $\sim$500 h of data ($\sim$700 h were acquired).  
A rapid shift to "remote" operations largely enabled a full-scale observing program 
in 2020-21, but it required pausing the bright-moon observing program 
($\sim$1000 h acquired; only $\sim$40 h bright-moon data); this will resume in Fall 2021. 
Excluding the pandemic-affected seasons, VERITAS averages $\sim$930 h / season
of good-weather observations during ``dark time'' (moon
illumination $<$30\%), and $\sim$200 h / season during periods of ``bright
moonlight'' (i.e. $>$30\% illumination). The bright-moon data has similar sensitivity to
dark-time observations, but has higher threshold (e.g. 250
GeV) \cite{BrightMoon_paper}.  

AGN comprise 63\% of the VERITAS source
catalog (shown in Figure~\ref{VERITAS_catalog}), and their observations are the dominant 
component of VERITAS data taking ($\sim$50\%).  As of July 2021, AGN data comprise 
a total of $\sim$5,900 h ($\sim$420 h per year) of good-weather dark time, $\sim$1,100 h ($\sim$140 h per year) of
good-weather, bright-moon time, and $\sim$1,800 h poor-weather (filler) observations. 
Blazars are the dominant component of the AGN program, and the 
good-weather dark time is historically split  $\sim$90\% to blazars, primarily BL Lac objects, 
and $\sim$10\% to radio-galaxies. From 2019$-$21, VERITAS acquired $\sim$670 h of good-weather 
dark time on blazars and $\sim$100 h of good-weather dark time on radio galaxies, showing a larger
emphasis on radio galaxies.  There were a further $\sim$90 h of good-weather, bright-moon time 
acquired on blazars.  After initially using 35$-$45\% of the bright-moon
time for blazar discoveries, almost all  of this time ($\sim$90\% from 2019-21) is now used for 
observing hard-spectrum BL Lac objects and searching for VHE AGN flares from targets not
in the regular VERITAS AGN monitoring program. The filler data is also used to search for AGN flares.
These supplemental flare monitoring programs have found several events not otherwise identified.

\begin{figure*}[t]
   \centerline{{\includegraphics[width=1.75in]{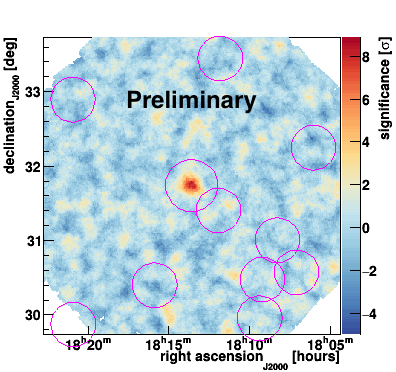} }
              \hfil
              {\includegraphics[width=4.0in]{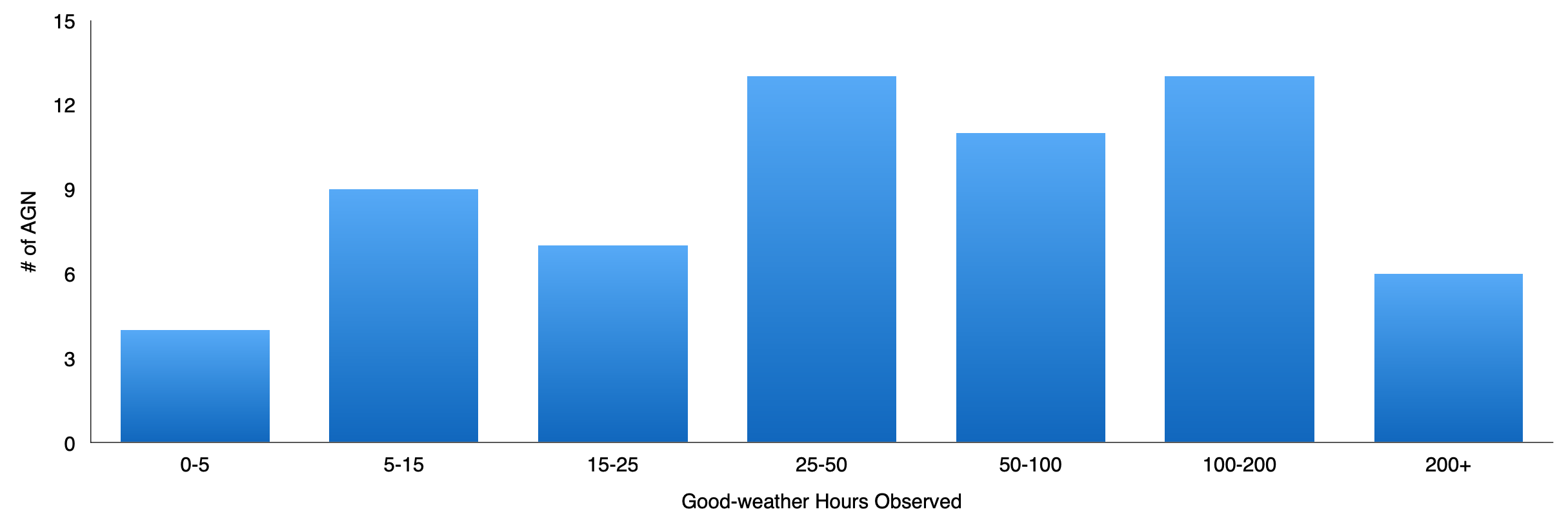} }
             }
   \caption{{\footnotesize (Left) The preliminary sky map of the significance measured from
     the direction of B2\,1811+31. (Right) Histogram of the total VERITAS exposure for every Northern VHE AGN; $\sim$50\%
     already have more than 50 h.}}
   \label{results_panel1}
\vspace{-0.2cm}
 \end{figure*}

Much of the AGN observing program is based on regular monitoring of 
selected objects in the Northern VHE catalog.  The depth and cadence of
these observations is based on a variety of criteria and scientific goals.
These exposures aim to both to self-identify VHE flaring episodes for immediate / intense target-of-opportunity (ToO) 
follow-up including MWL partners, while simultaneously building deeper, legacy 
exposures on particularly interesting objects. Figure~\ref{results_panel1} shows the distribution of exposures
already achieved by VERITAS for each Northern AGN. The monitoring cadence 
ranges from daily to weekly, while the minimum sample duration will detect $\sim$10\% Crab
flux. While the general goal of the program is to identify long-lasting, flare states for 
intense campaigns, our experience shows that we are also
able to fortuitously catch short-duration, bright flares (e.g. BL\,Lac in 2017
\cite{BLLAC_2017}). Contemporaneous MWL data (e.g. Swift) are timed with the
VERITAS monitoring observations to both assist with triggering and to enable long-term modeling.
For the 2020-21 season, VERITAS monitored 23 VHE AGN (37\% of the Northern population).

While a major focus of the AGN program is performing deep / timely measurements of
known VHE sources, $\sim$35\% of the 2019-21 data were devoted to the 
discovery of new VHE AGN. This included regular observations of
targets from a list of selected candidates, and ToO observations triggered by our partners.
Our discovery candidates include several AGN with a weak excess ($>$3$\sigma$) in
large, archival VERITAS exposures; these excesses continue to grow in limited annual exposures.
In addition, we continue a program to observe all targets from a comprehensive list of Northern discovery candidates
for at least 5 h. Only 11 objects require further observation, and most have at least some exposure.
The target list includes all the X-ray bright HBLs in the 2WHSP catalog 
(``TeV Figure of Merit'' $>$ 1.0; \cite{2WHSP}), all the hardest
AGN in the {\it Fermi}-LAT $>$50 GeV catalog ($\Gamma_{2FHL} < 2.8$; \cite{2FHL_Catalog}, 
all nearby ($z < 0.3$) LBLs from the MOJAVE sample with relatively high maximum apparent jet speed \cite{Lister}, 
and all targets from prior comprehensive efforts \cite{Benbow09,Benbow11}. 

ToO observations are the highest priority of the VERITAS program.
These data comprised 12\% and 15\% of the blazar dark-time in the 2019-20 and
2020-21 seasons, respectively.  Approximately 70\% of these data were 
follow-up observations of flares in known VHE blazars, with the remainder
including attempts to discover and/or follow-up on new VHE blazars.

\section{Recent Highlights}

{\bf B2\,1811+31} is an IBL at redshift $z = 0.117$ that showed an elevated MeV-GeV flux ($\sim$11x brighter) and
harder gamma-ray spectrum ($\Gamma_{\mathrm LAT} \sim 1.4$ vs. 2.1) during \emph{Fermi}-LAT observations in Oct.
2020 (ATel \#14060). Following the discovery of VHE emission by MAGIC (Oct. 4-10; ATel \#14090), 
and reports of enhanced optical activity (ATel \#14103), VERITAS observed the blazar from Oct. 15-19, 2020.
A preliminary analysis of these data ($\sim$5 h) yields a strong detection ($\sim$8 standard deviations, $\sigma$) and a soft
photon spectrum ($\Gamma$ = 4.1 $\pm$ 0.5). A sky map of the significance observed 
near B2\,1811+31 is shown in Figure~\ref{results_panel1}.  The VERITAS light curve
is consistent with a constant flux F($>$ 250 GeV) = $(1.10 \pm 0.18_{\mathrm stat}) \times 10^{-11}$ cm$^{-2}$ s$^{-1}$.
This is approximately 6\% of the Crab Nebula flux (Crab) above the same threshold, and is similar to
the flux reported by MAGIC.

\begin{figure*}[t]
   \centerline{{\includegraphics[width=3.25in]{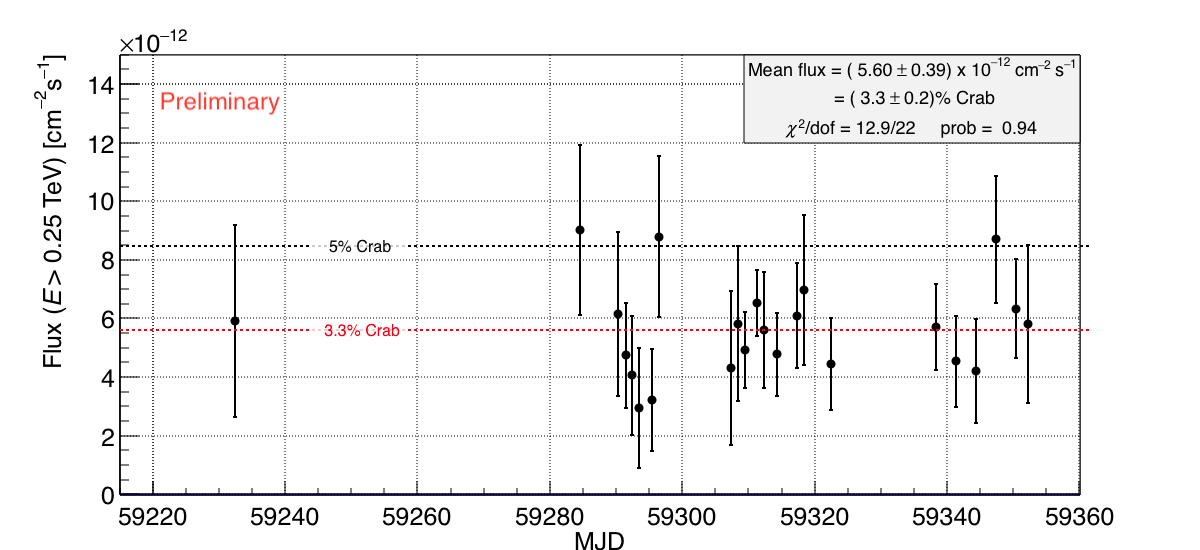} }
              \hfil
              {\includegraphics[width=2.25in]{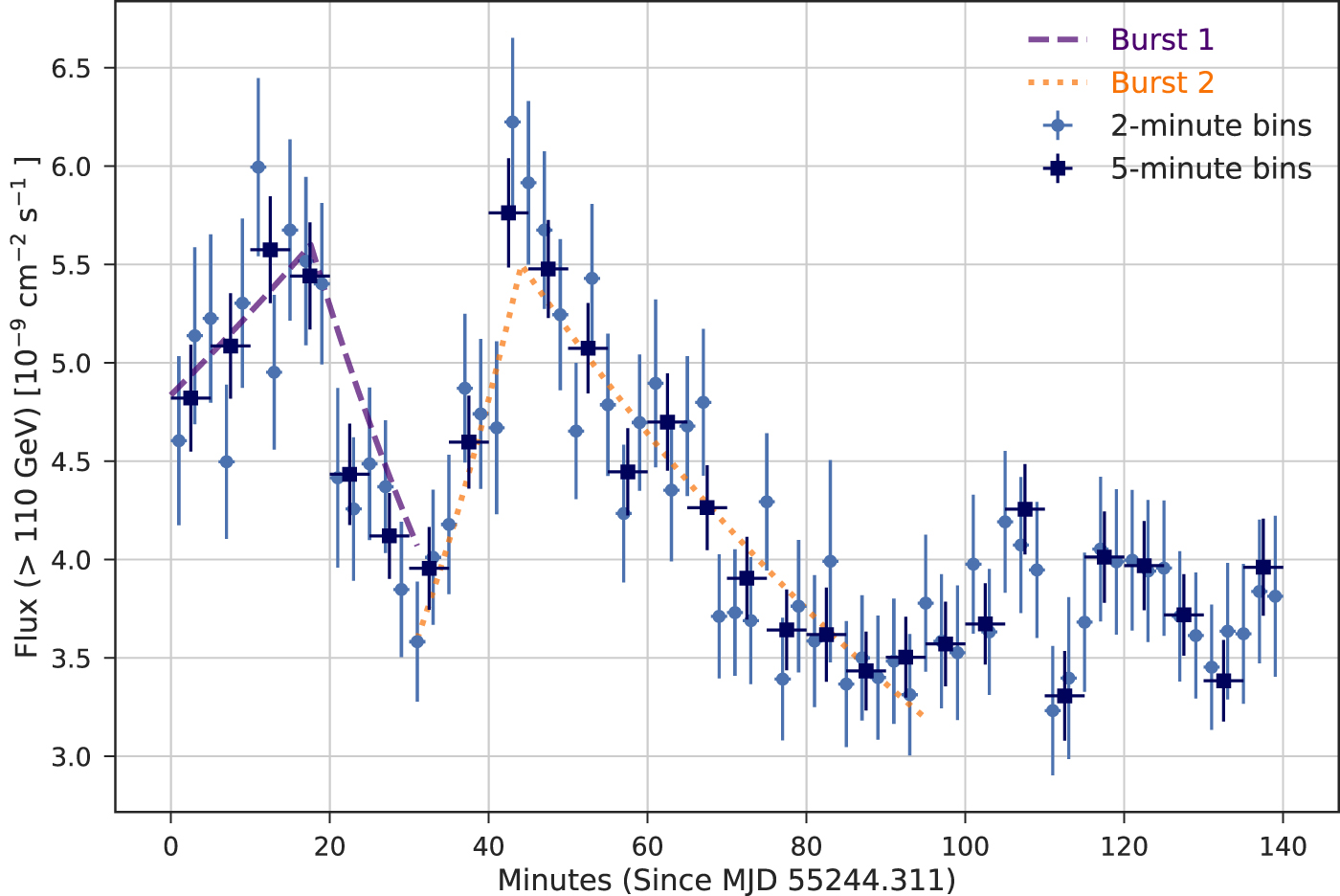} }
             }
   \caption{{\footnotesize  (Left) The preliminary nightly light curve from VERITAS
     observations of H\,1426+428 in 2021. (Right) The VERITAS light curve from Mrk 421 on Feb. 17, 2010 \cite{VERITAS_421_flare}. 
   The dashed lines are exponential fits to the two bursts (2-min-bins).
   The resulting timescales are 84 and 22 min for doubling (rising), and 28 and 65 min for halving (decay).}}
   \label{results_panel2}
\vspace{-0.2cm}
 \end{figure*}

{\bf H\,1426+428} ($z = 0.129$) is an extreme HBL with synchrotron peak located at 10$^{18.1}$ Hz \cite{2WHSP}.  
It was routinely detected before 2002 with VHE fluxes between $\sim$5\% and $\sim$20\%
Crab. However, since this time the reported VHE fluxes are significantly less ($<$2-3\% Crab).
H\,1426+428 was observed with VERITAS in almost every season and a $\sim$200 h exposure exists.
A preliminary analysis of $\sim$82 h of good-quality data from 2008-2016 yields a 13$\sigma$ detection.  
Although MWL observations show variations (e.g., the Swift XRT count rate varies by a factor of 3), 
the VHE flux F($>$250 GeV) = (1.9 $\pm$ 0.2)\% Crab shows no signs of variability.
The preliminary VHE spectrum can be described 
by a power law with $\Gamma = 2.8 \pm 0.1$.  During 2021 monitoring observations,
H\,1426+428 was found to be in a bright VHE state. This triggered an intense VERITAS and MWL campaign
including Swift and NuStar.  Overall, $\sim$45 h of VERITAS data were acquired.  A preliminary analysis
of these data yields a $\sim$19$\sigma$ detection and a time-average spectrum ($\Gamma \sim 2.6$) consistent
with the long-term measurement.  The preliminary flux, 
F($>$250 GeV) =  $(5.6 \pm 0.4_{\mathrm stat}) \times 10^{-12}$ cm$^{-2}$ s$^{-1}$, 
or (3.3 $\pm$ 0.2)\% Crab, is steady throughout 2021 but higher than the 2008-16 average.
The preliminary light curve from the 2021 observations is shown in Figure~\ref{results_panel2}.
The steady VERITAS flux contrasts with the Swift-XRT monitoring which indicates a high count rate
and significant variability.

{\bf 3C\,264}  is an FR-I type radio galaxy at $z =0.0216$.  It was observed with VERITAS for 
57 hr from 2017-19, resulting in the 7.8$\sigma$ discovery of VHE emission \cite{VERITAS_3C264}.
Its VHE flux is variable on monthly timescales and was elevated in 2018. 
The hard-spectrum ($\Gamma \sim 2.20$) VHE emission during 2018 has
F($>$315 GeV) = ($7.6\pm 1.2_{\mathrm stat} \pm 2.3_{\mathrm syst})\times 10^{-13}$ cm$^{-2}$ s$^{-1}$ (0.7\% Crab). 
The elevated VHE state was thought to be similar to those observed 
from the analogous source, M\,87, and extensive contemporaneous MWL data 
were acquired in 2018, including high-resolution imaging data (e.g., Chandra, HST, VLA, VLBA). However, there was 
no clearly identifiable source of the elevated flux. The 3C 264 SED
is unusual for a radio galaxy; its synchrotron peak near the X-rays has relatively high frequency. 
However, many aspects of the SED can be qualitatively reproduced with an SSC model using parameters typical
of BL Lacs. Comparing the SEDs of 3C\,264 and M\,87 shows differences that are plausibly
explained by 3C\,264 being oriented closer to the line of sight.

{\bf Mrk\,421} exhibited an extraordinary flaring episode in February 2010 \cite{VERITAS_421_flare}. 
The VHE flux observed by VERITAS from this HBL during an intense MWL campaign 
reached 27 Crab, the highest ever observed from an AGN.  The light-curve from the brightest
night is shown in Figure~\ref{results_panel2} and enables detailed cross-correlation analyses.  
Limits on the Doppler factor ($\delta \gtrapprox 33$)
and the size of the emission region ($R_{B} / \delta \lessapprox 3.8 \times 10^{13}$ cm) 
are obtained from the fast variability. A lag (25-55 minutes) between the VHE and optical bands
is seen (3$\sigma$) for the first time on short time scales. The VHE and X-ray fluxes show a wide range of behavior, including linear and quadratic correlations, as well as anti-correlations. The MWL data are difficult to explain using a single-zone SSC model.

{\bf TXS\,0506+056} is an important object for multi-messenger astronomy (see \cite{0506_VERITAS} and references
therein). It was observed to have an
elevated VHE and MeV-GeV gamma-ray flux in spatial and temporal coincidence ($\sim$3$\sigma$) with 
the IceCube high energy neutrino event IC170922A.
This could indicate that blazar jets accelerate cosmic rays to
at least several PeV, and are hence a source of VHE cosmic rays.
The association was also used to herald the birth of neutrino astronomy.
The initial VERITAS follow-up observations of the neutrino/blazar 
($\sim$35 h from Sept. 2017 to Feb. 2018) led to a soft-spectrum ($\Gamma \sim 4.8$), VHE
detection (5.8$\sigma$),
albeit at lower VHE flux than detected by MAGIC.  VERITAS has since carried out deep
observations of TXS\,0506+056 and an associated MWL campaign \cite{Jin_ICRC21}.  A weak excess (3.4$\sigma$)
is observed in $\sim$61 h collected from Oct. 2018 to Feb. 2021.
Interpreting this excess as a detection, it corresponds to F($>$190 GeV) $\sim 0.5$\% Crab.
This is slightly lower than, but statistically consistent with, the prior VERITAS flux ($\sim$0.7\% Crab).

{\bf The TeV luminosity function} of HBLs is important because these AGN
dominate the extragalactic VHE sky and hence the total cosmic VHE radiation.
Its measurement (i.e. the number of HBLs per unit volume per unit luminosity)
is key to understanding HBL properties, their relationship with other sources, 
and their contributions to unresolved radiation fields. This measurement is challenging 
due to observational biases, but enables 
studies of hadronic/neutrino production in jets, the IGMF, and AGN evolution.
A program \cite{Errando_ICRC21} was designed to minimize these biases by selecting 
36 HBLs from the 3WHSP catalog, and measuring their VHE fluxes at times not weighted towards high-fluxes. 
These VERITAS observations  are complete and leverage $\sim$1800 h of archival data and $\sim$150 h of
2019-21 data. Each target has at least 8 h of exposure ($\sim$1\% Crab sensitivity).

{\bf FSRQs:} are generally detected at VHE energies during flaring states. VERITAS has detected
3 FRSQs and most of its FSRQ observations are ToO based. In 2020, VERITAS began the first
systematic search for VHE emission from FSRQs. Twelve objects were selected for
at least 8 hours of unbiased (non-ToO) observations based on the 3FHL catalog and/or
prior VHE detection. The data provide a sensitivity to fluxes of $\sim$1\% Crab.
Upper limits from the first 4 FSRQ (GB6\,J0043+3426, S3\,0218+35, PKS\,0736+17 and 3C\,454.3)
are described in \cite{Patel_ICRC21}.  When complete, the survey will
provide the first constraints on the duty cycle of VHE emission from FSRQs.

\section{Conclusion}

As of July 2021, the VERITAS collaboration has acquired $\sim$7,000 good-weather hours targeted on AGN.  
Since \emph{ICRC 2019}, the array was used to acquire $\sim$860 h of these observations. 
Unfortunately, each of the past two seasons were affected by the global
pandemic, with yields reduced by a $\sim$4-month observatory
closure in 2019-20, and a temporary suspension of bright-moon observing in 2020-21.
Overall, the 2019-21 AGN yield was $\sim$25\% below the two-year average.

The VERITAS AGN program continued to focus on deep, regular VHE and MWL monitoring
of known VHE AGN, and immediate and intense ToO follow-up of interesting
flaring events. We also maintained a robust discovery program, with 
$\sim$35\% of our most recent observations having a discovery focus.
Highlights from our recent AGN observations and publications include the
observation of flares from B2\,1811+31, H\,1426+428, 3C\,264 and Mrk\,421.

The VERITAS array is now $\sim$14 years old and continues to run well.
Indeed, the past four seasons each rank among the five best for various technical
performance benchmarks (e.g., fraction of data with all telescopes operational). 
The collaboration plans to operate VERITAS through at least 2025, and will
continue prioritizing AGN observations. Given the array's strong technical performance, 
we expect the long VERITAS tradition of producing exciting AGN results to continue.

\vspace{0.2cm}

\footnotesize{This research is supported by grants from the U.S. Department of Energy Office of Science, the U.S. National Science Foundation and the Smithsonian Institution, by NSERC in Canada, and by the Helmholtz Association in Germany. This research used resources provided by the Open Science Grid, which is supported by the National Science Foundation and the U.S. Department of Energy's Office of Science, and resources of the National Energy Research Scientific Computing Center (NERSC), a U.S. Department of Energy Office of Science User Facility operated under Contract No. DE-AC02-05CH11231. We acknowledge the excellent work of the technical support staff at the Fred Lawrence Whipple Observatory and at the collaborating institutions in the construction and operation of the instrument.}

%
%
%

\clearpage \section*{Full Authors List: \Coll\ Collaboration}

\scriptsize
\noindent
C.~B.~Adams$^{1}$,
A.~Archer$^{2}$,
W.~Benbow$^{3}$,
A.~Brill$^{1}$,
J.~H.~Buckley$^{4}$,
M.~Capasso$^{5}$,
J.~L.~Christiansen$^{6}$,
A.~J.~Chromey$^{7}$, 
M.~Errando$^{4}$,
A.~Falcone$^{8}$,
K.~A.~Farrell$^{9}$,
Q.~Feng$^{5}$,
G.~M.~Foote$^{10}$,
L.~Fortson$^{11}$,
A.~Furniss$^{12}$,
A.~Gent$^{13}$,
G.~H.~Gillanders$^{14}$,
C.~Giuri$^{15}$,
O.~Gueta$^{15}$,
D.~Hanna$^{16}$,
O.~Hervet$^{17}$,
J.~Holder$^{10}$,
B.~Hona$^{18}$,
T.~B.~Humensky$^{1}$,
W.~Jin$^{19}$,
P.~Kaaret$^{20}$,
M.~Kertzman$^{2}$,
T.~K.~Kleiner$^{15}$,
S.~Kumar$^{16}$,
M.~J.~Lang$^{14}$,
M.~Lundy$^{16}$,
G.~Maier$^{15}$,
C.~E~McGrath$^{9}$,
P.~Moriarty$^{14}$,
R.~Mukherjee$^{5}$,
D.~Nieto$^{21}$,
M.~Nievas-Rosillo$^{15}$,
S.~O'Brien$^{16}$,
R.~A.~Ong$^{22}$,
A.~N.~Otte$^{13}$,
S.~R. Patel$^{15}$,
S.~Patel$^{20}$,
K.~Pfrang$^{15}$,
M.~Pohl$^{23,15}$,
R.~R.~Prado$^{15}$,
E.~Pueschel$^{15}$,
J.~Quinn$^{9}$,
K.~Ragan$^{16}$,
P.~T.~Reynolds$^{24}$,
D.~Ribeiro$^{1}$,
E.~Roache$^{3}$,
J.~L.~Ryan$^{22}$,
I.~Sadeh$^{15}$,
M.~Santander$^{19}$,
G.~H.~Sembroski$^{25}$,
R.~Shang$^{22}$,
D.~Tak$^{15}$,
V.~V.~Vassiliev$^{22}$,
A.~Weinstein$^{7}$,
D.~A.~Williams$^{17}$,
and 
T.~J.~Williamson$^{10}$\\
\noindent
$^1${Physics Department, Columbia University, New York, NY 10027, USA}
$^{2}${Department of Physics and Astronomy, DePauw University, Greencastle, IN 46135-0037, USA}
$^3${Center for Astrophysics $|$ Harvard \& Smithsonian, Cambridge, MA 02138, USA}
$^4${Department of Physics, Washington University, St. Louis, MO 63130, USA}
$^5${Department of Physics and Astronomy, Barnard College, Columbia University, NY 10027, USA}
$^6${Physics Department, California Polytechnic State University, San Luis Obispo, CA 94307, USA} 
$^7${Department of Physics and Astronomy, Iowa State University, Ames, IA 50011, USA}
$^8${Department of Astronomy and Astrophysics, 525 Davey Lab, Pennsylvania State University, University Park, PA 16802, USA}
$^9${School of Physics, University College Dublin, Belfield, Dublin 4, Ireland}
$^10${Department of Physics and Astronomy and the Bartol Research Institute, University of Delaware, Newark, DE 19716, USA}
$^{11}${School of Physics and Astronomy, University of Minnesota, Minneapolis, MN 55455, USA}
$^{12}${Department of Physics, California State University - East Bay, Hayward, CA 94542, USA}
$^{13}${School of Physics and Center for Relativistic Astrophysics, Georgia Institute of Technology, 837 State Street NW, Atlanta, GA 30332-0430}
$^{14}${School of Physics, National University of Ireland Galway, University Road, Galway, Ireland}
$^{15}${DESY, Platanenallee 6, 15738 Zeuthen, Germany}
$^{16}${Physics Department, McGill University, Montreal, QC H3A 2T8, Canada}
$^{17}${Santa Cruz Institute for Particle Physics and Department of Physics, University of California, Santa Cruz, CA 95064, USA}
$^{18}${Department of Physics and Astronomy, University of Utah, Salt Lake City, UT 84112, USA}
$^{19}${Department of Physics and Astronomy, University of Alabama, Tuscaloosa, AL 35487, USA}
$^{20}${Department of Physics and Astronomy, University of Iowa, Van Allen Hall, Iowa City, IA 52242, USA}
$^{21}${Institute of Particle and Cosmos Physics, Universidad Complutense de Madrid, 28040 Madrid, Spain}
$^{22}${Department of Physics and Astronomy, University of California, Los Angeles, CA 90095, USA}
$^{23}${Institute of Physics and Astronomy, University of Potsdam, 14476 Potsdam-Golm, Germany}
$^{24}${Department of Physical Sciences, Munster Technological University, Bishopstown, Cork, T12 P928, Ireland}
$^{25}${Department of Physics and Astronomy, Purdue University, West Lafayette, IN 47907, USA}

\end{document}